\documentclass{emulateapj}

\newcommand{\kms}{\, {\rm km\, s}$^{-1}$}

\begin{document}
\title{Correlations between \ion{O}{6} Absorbers and Galaxies at Low Redshift}

\author{
Rajib Ganguly\altaffilmark{1},
Renyue Cen\altaffilmark{2},
Taotao Fang\altaffilmark{3} and
Kenneth Sembach\altaffilmark{4}}

\begin{abstract}

We investigate the relationship between galaxies and metal-line
absorption systems in a large-scale cosmological simulation with
galaxy formation.  Our detailed treatment of metal enrichment and
non-equilibrium calculation of oxygen species allow us, for the
first time, to carry out quantitative calculations of the
cross-correlations between galaxies and \ion{O}{6} absorbers. We
find the following: (1) The cross-correlation strength depends
weakly on the absorption strength but strongly on the luminosity of
the galaxy. (2) The correlation distance increases monotonically
with luminosity from $\sim 0.5-1h^{-1}$Mpc for $0.1L_*$ galaxies to
$\sim 3-5h^{-1}$Mpc for $L_*$ galaxies. (3) The correlation distance
has a complicated dependence on absorber strength, with a
luminosity-dependent peak. (4) Only 15\% of \ion{O}{6} absorbers lie
near $\ge L_\mathrm{z,*}$\ galaxies. The remaining 85\%, then, must
arise ``near'' lower-luminosity galaxies, though, the positions of
those galaxies is not well-correlated with the absorbers. This may
point to pollution of intergalactic gas predominantly by smaller
galaxies. (5) There is a subtle trend that for $\gtrsim
0.5L_\mathrm{z,*}$ galaxies, there is a positive correlation between
absorber strength and galaxy luminosity in the sense that stronger
absorbers have a slightly higher probability of finding such a large
galaxy at a given projection distance. For less luminous galaxies,
there seems to be a negative correlation between luminosity and
absorber strength.
\end{abstract}


\altaffiltext{1}{Department of Physics \& Astronomy, University of
Wyoming, 1000 E. University Ave., Laramie, WY 82071, email:
ganguly@uwyo.edu}

\altaffiltext{2}{Department of Astrophysical Sciences, Princeton
University, Peyton Hall, Ivy Lane, Princeton, NJ 08544, email:
cen@astro.princeton.edu}

\altaffiltext{3}{Department of Physics \& Astronomy, University of
California, 4129 Frederick Reines Hall, Irvine, CA 92697, email:
fangt@uci.edu}

\altaffiltext{4}{The Space Telescope Science Institute, 3700 San
Martin Drive, Baltimore, MD 21218, email: sembach@stsci.edu}

\section{Introduction}

Cosmological hydrodynamic simulations have shown that most of the
so-called missing baryons \citep*{fhp98} may be in a filamentary
network of Warm-Hot Intergalactic Medium \citep[WHIM;][]{co99,d01}.
Visual inspection of simulations suggests that the WHIM is spatially
correlated with galaxies. This is consistent with the observed
large-scale structure of galaxies, as well as the physical
expectation that both galaxies and intergalactic medium (IGM) are
subject to the dominant gravitational force of dark matter which
tends to lead to such large-scale structures \citep{z70}.

Moreover, the WHIM may provide a primary conduit for matter and
energy exchanges between galaxies and the IGM. Thus, a detailed
understanding of the WHIM may shed useful light on galaxy formation
\citep*[e.g., ][]{cno05}. The \ion{O}{6} $\lambda \lambda$1032, 1038
absorption line doublet in the spectra of low-redshift QSOs provides
a valuable probe of the WHIM \citep*[e.g.,][]{l07,sd07,t08}. In this
{\it Letter} we use cosmological hydrodynamic simulations to make
predictions of the correlations between \ion{O}{6} absorbers and
galaxies, which may be used for detailed comparisons with upcoming
{\it Hubble Space Telescope Cosmic Origins Spectrograph}
observations.

\section{Simulation and Construction of Absorber and Galaxy
Catalogs}

We use the simulation from \citet{co06} and \citet{cf06},
%which includes feedback from galactic stellar winds and follows the
%non-equilibrium ionization in a self-consistent manner,
to make predictions of the relationship between \ion{O}{6} WHIM
absorption and the presence of galaxies. This cold dark matter
simulation assumes: $\Omega_M = 0.31$, $\Omega_b = 0.048$,
$\Omega_\Lambda=0.69$, $\sigma_8 = 0.89$, $H_0 = 100 h=69$\kms
Mpc$^{-1}$, $n_s = 0.97$, co-moving box size $85 h^{-1}$\,Mpc and
$1024^3$\ cells. %The cell size is 83~h$^{-1}$\,kpc.
The simulation follows star formation using a physically-motivated
prescription and includes feedback processes from star formation to
the IGM in the form of UV radiation and galactic superwinds carrying
energy and metal-enriched gas \citep{co06}. In star-formation sites,
``star particles'' are produced at each time step, which typically
have a mass of $10^6$\,M$_\odot$. Galaxies are identified post facto
using the HOP grouping scheme \citep[][see Nagamine et~al. 2001 for
details]{Eisenstein98} on these particles. This grouping scheme
provides a catalog of galaxies containing 3D positions, peculiar
velocities, stellar masses and ages. The catalog consists of 33,887
galaxies. Stellar population synthesis models from \citet{bc03} are
used to compute luminosities in all SDSS bands. For the purposes of
this paper, we focus on the Sloan z-band, which is centered at
9100\,\AA, and has a width of 1200\,\AA\ \citep{sdss}. For galaxies,
the flux in this band is most tightly correlated (of all the Sloan
bands) with the total stellar mass \citep[e.g.,][]{kauffmann03}. We
find that the luminosities of field galaxies in this simulation
follow a Schechter function down to $\sim10^{-3.5} L_\mathrm{z,*}$\
($L_\mathrm{z,*} = 3 \times 10^{11}\,L_\odot$, Ganguly et al. 2008,
in preparation).

We extract a {$100 \times 100$} grid of sight-lines uniformly
separated by $850 h^{-1}$\,kpc. For each sight-line, a synthetic
spectrum of \ion{O}{6} $\lambda$1032 absorption is generated, taking
into account effects due to peculiar velocities and thermal
broadening. For each sight-line, we decompose the spectrum into
individual Gaussian-broadened components with an algorithm similar
to {\sc autovp} \citep{autovp}. Finally, we associate grouped
components into systems; this is important since, physically, it
does not serve our purpose to treat individual components separately
when comparing to the locations of galaxies. [A single physical
system like a galactic disk/halo is often composed of multiple
components. This is a result of complicated velocity structures and
the clumpiness of gas. Historically, older surveys did not have the
spectral dispersion needed to resolve individual components.
Therefore, it is desirable to re-group these components back to
physically independent systems.] We accomplish this by computing the
one-dimensional two-point correlation function of components and
identify a characteristic velocity scale. We find that, on small
scales, absorption-line components are correlated out to a velocity
separation of $\sim\!\!\pm300$\,\kms, which we adopt as the
characteristic velocity interval to identify systems. For each
system, we record the integrated flux-weighted centroid redshift,
\ion{O}{6} column density, and $\lambda$1032 rest-frame equivalent
width. A total of $\sim$180,000 components were found to be grouped
into $\sim$21,000 systems with $W_\lambda(\lambda1032) \ge 1$\,m\AA.

\begin{figure}[t]
\epsscale{1.2}
\plotone{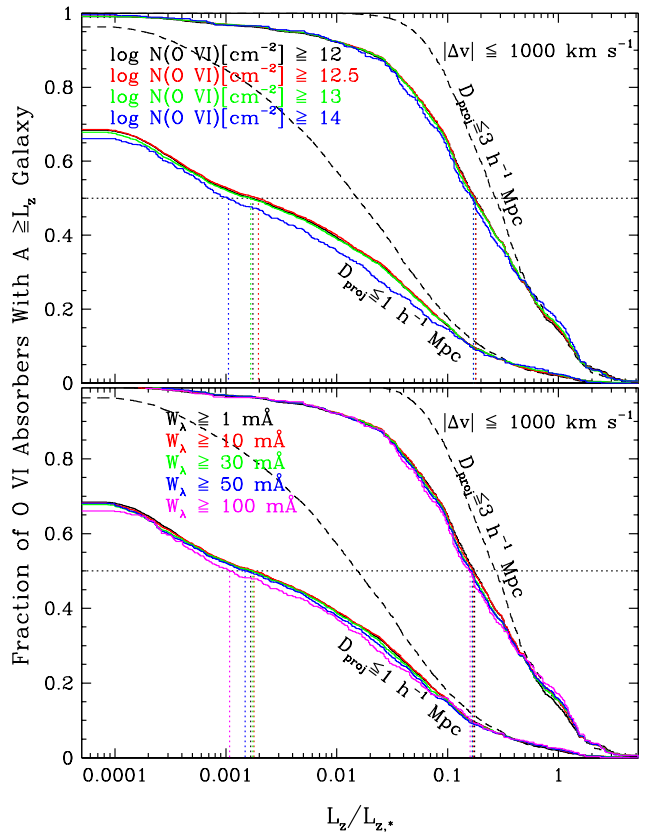} \caption{We show the cumulative
fraction of \ion{O}{6} absorbers that have galaxy at least as
luminous as $L_\mathrm{z}/L_\mathrm{z,*}$\ within a cylindrical
volume centered on the absorber. We show results for cylinders of
two different radii, and a velocity depth of 2000\kms. In the top
panel, we show different cuts in the integrated \ion{O}{6} column
density. In the bottom panel, we show cuts in $\lambda$1032
equivalent width. In each panel, a horizontal line is drawn where
50\% of the absorbers are accounted for. A vertical line is drawn
where each of the curves crosses this fiducial. Black dashed lines
show the results for uncorrelated absorber and galaxy positions.}
\label{fig:1}
\end{figure}

\section{Analysis \&\ Results}

We consider the probability that an \ion{O}{6} absorber of a given
equivalent width or column density lies within some distance of a
galaxy with a certain luminosity. We take three approaches to
address this question, but we must first tackle the problem of
assigning galaxies to \ion{O}{6} absorbers. Due to peculiar velocity
effects, the exact 3D separation between an \ion{O}{6} absorber and
a galaxy cannot be precisely known, although the projected distance
in the sky plane can be directly measured. Thus, we simply limit
absorber-galaxy associations to within the physically motivated
line-of-sight separation of $1000$\kms, corresponding approximately
to the velocity dispersion of clusters of galaxies.

Approach \#1: For each absorber, we find the most luminous galaxy
within a projected distance of 1h$^{-1}$\,Mpc or 3h$^{-1}$\,Mpc (and
within the aforementioned velocity separation).  Figure~\ref{fig:1}
shows the cumulative fraction of \ion{O}{6} absorbers that have a
galaxy of luminosity $>L_\mathrm{z}$ within this cylindrical
volume. We show this distribution for \ion{O}{6} absorbers of
different integrated column density cuts (upper panel) or
$\lambda$1032 equivalent width cuts (lower panel). For comparison, we
also do the same exercise for a sample of galaxies placed randomly in
the box, but with the same luminosity function and the same total
number of galaxies.

\begin{figure}[t]
\epsscale{1.2}
\plotone{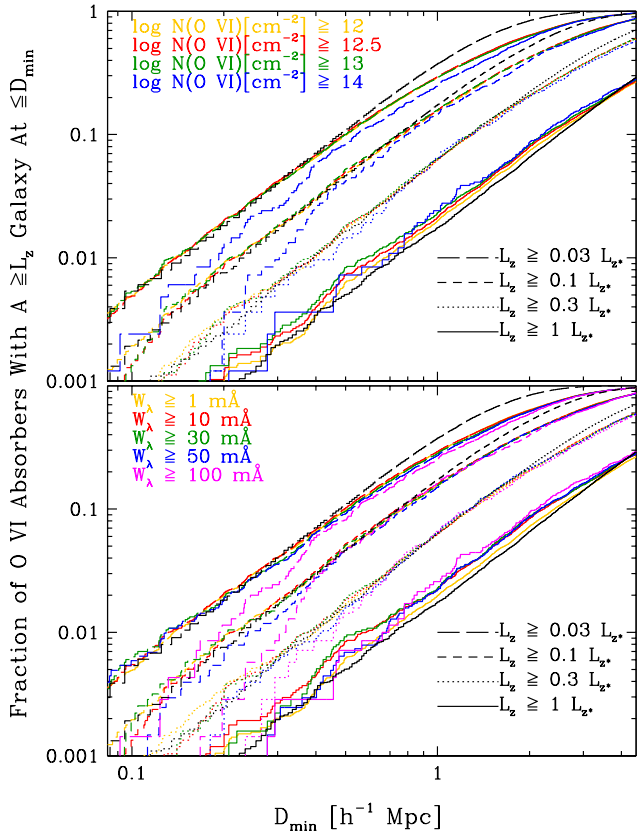} \caption{We show the cumulative
fraction of \ion{O}{6} absorbers as a function of projected distance
to the closest galaxy with luminosity $\ge L_\mathrm{z}$. The
absolute velocity separation is required to be $|\Delta v| \le
1000$\kms. We show different cuts in \ion{O}{6} column density (top
panel) and $\lambda$1032 equivalent width (bottom panel). In both
panels, four families of curves are shown corresponding to different
luminosity cuts: $L_\mathrm{z}/L_\mathrm{z,*} \ge 0.03, 0.1, 0.3$\
and 1. Black lines show the results for uncorrelated absorber and
galaxy positions. Note that vertical cuts at $D_\mathrm{min}=1,
3$h$^{-1}$\,Mpc plotted against luminosity reproduce
Figure~\ref{fig:1}.} \label{fig:2}
\end{figure}

Approach \#2: We relax the projected distance requirement,
and find the closest galaxy with luminosity $\ge f L_\mathrm{z,*}$\
(with four different $f=0.03,0.1,0.3,1$). Figure~\ref{fig:2} shows
the cumulative fraction of \ion{O}{6} absorbers as a function of
projected distance. As with Figure~\ref{fig:1}, we show different
column density (upper panel) and equivalent width cuts (lower
panel). In addition, we show the different cuts in galaxy
luminosity, and we repeat the exercise with a sample of randomly
placed galaxies. Note that Figures~\ref{fig:1} and \ref{fig:2} are
complementary. A vertical cut in Figure~\ref{fig:2} at
1h$^{-1}$\,Mpc or 3h$^{-1}$\,Mpc, with the cumulative fraction of
absorbers plotted against luminosity reproduces Figure~\ref{fig:1}.

Approach \#3; For each absorber, we determine the number of galaxies
that lie within the 3h$^{-1}$\,Mpc radius, 2000\,\kms\ deep
cylindrical volume. In Table~\ref{tab:ngal}, we list the fraction of
these absorbers that have more than n galaxies ($n=0,1,...$) in that
volume. We do this for two subsamples of absorbers, those with
$\lambda1032$\ equivalent width larger than 30\,m\AA\ (typical
detection limit for HST/STIS and FUSE spectra) and 10\,m\AA
(expected limit for HST/COS observations). In addition to the
equivalent width cuts in the absorber sample, we also make
luminosity cuts in the galaxy sample. Current catalogs are typically
able to reach galaxy luminosities of 0.1$L_*$\ for statistically
interesting volumes \citep[e.g.,][]{s06}.

\begin{deluxetable}{lllllll}
\tablewidth{0pc}
\tablecaption{\ion{O}{6} Absorbers vs. Galaxies}

\tablehead {
%& \multicolumn{6}{c}{$f = L_\mathrm{z}/L_\mathrm{z,*}$} \\
%& \multicolumn{6}{c}{\hrulefill} \\
\colhead{n}  & \colhead{$f=1$} & \colhead{0.3} & \colhead{0.1} & \colhead{0.03} & \colhead{0.01} & \colhead{0}}
\startdata
\multicolumn{7}{c}{$W_\lambda(\lambda1032)\ge$10\,m\AA, $\sim$6500\ absorbers} \\ \hline  \\[-7pt]
0  & 0.15  &  0.36 &   0.56 &   0.74 &   0.84 &   0.99 \\
1  & 0.04  &  0.17 &   0.40 &   0.62 &   0.75 &   0.98 \\
2  & 0.01  &  0.08 &   0.28 &   0.52 &   0.67 &   0.96 \\
3  & 0.00  &  0.04 &   0.20 &   0.43 &   0.60 &   0.94 \\
4  & 0.00  &  0.02 &   0.14 &   0.36 &   0.52 &   0.92 \\
5  & 0.00  &  0.02 &   0.10 &   0.30 &   0.47 &   0.89 \\
10 & 0.00  &  0.00 &   0.02 &   0.12 &   0.26 &   0.75 \\  \hline \\[-7pt]
\multicolumn{7}{c}{$W_\lambda(\lambda1032)\ge$30\,m\AA, $\sim$3000 absorbers} \\ \hline  \\[-7pt]
0  & 0.15  &  0.35 &   0.55 &   0.74 &   0.84 &   0.99 \\
1  & 0.05  &  0.17 &   0.39 &   0.61 &   0.75 &   0.98 \\
2  & 0.01  &  0.08 &   0.27 &   0.51 &   0.67 &   0.96 \\
3  & 0.01  &  0.05 &   0.19 &   0.42 &   0.59 &   0.94 \\
4  & 0.00  &  0.02 &   0.14 &   0.35 &   0.52 &   0.92 \\
5  & 0.00  &  0.02 &   0.10 &   0.29 &   0.46 &   0.89 \\
10 & 0.00  &  0.00 &   0.02 &   0.11 &   0.25 &   0.75
\enddata
\tablecomments{Column 1 denotes the number of galaxies, n, that lie
within a cylindrical volume of radius 3h$^{-1}$\,Mpc and depth
2000\,\kms\ centered on an \ion{O}{6} absorber. In the header row,
$f$\ denotes the galaxy luminosity in units of $L_\mathrm{z,*}$\
(i.e., $L_\mathrm{z} = f L_\mathrm{z,*}$). The numbers in columns
2-7 indicate the fraction of $\ge W_\lambda(\lambda1032)$\
\ion{O}{6} absorbers that have $>$n galaxies with luminosities
$\ge f L_\mathrm{z,*}$.} \label{tab:ngal}
\end{deluxetable}

\begin{deluxetable}{llllll}
\tablewidth{0pc} \tablecaption{\ion{O}{6} Absorber - Galaxy
Correlation Length}

\tablehead {
& \multicolumn{5}{c}{$W_\lambda$(\ion{O}{6} $\lambda1032$) [m\AA]} \\
& \multicolumn{5}{c}{\hrulefill} \\
\colhead{$L_\mathrm{z}/L_\mathrm{z,*}$} & \colhead{1} & \colhead{10} & \colhead{30} & \colhead{50} & \colhead{100}}
\startdata
1    & 3.23    & 4.15    & 4.20    & 4.18    & 4.92 \\
0.3  & 1.17    & 1.40    & 1.84    & 1.94    & 1.27 \\
0.1  & 0.59    & 0.82    & 0.74    & 0.57    & \nodata \\
0.03 & \nodata & \nodata & \nodata & \nodata & \nodata
\enddata
\tablecomments{Column 1 denotes the luminosity of galaxies in units of
$L_\mathrm{z,*} = 3 \times 10^{11}\,L_\odot$. The numbers in columns 2-6
indicate the distance in Mpc out to which $\ge W_\lambda($\ion{O}{6} $\lambda1032)$\
absorbers are correlated with $\ge L_\mathrm{z}$\ galaxies. Galaxies are required to lie
within 1000\,\kms\ of the absorber.}
\label{tab:distlum}
\end{deluxetable}

Taking Figures~\ref{fig:1} and \ref{fig:2} together, we first
address the question of which galaxies, and on what scales, are
correlated (if any) with \ion{O}{6} WHIM absorption. In
Figure~\ref{fig:1}, at luminosities fainter than $0.3-0.5
L_\mathrm{z,*}$, the curves lie below the equivalent curves for
randomly-placed galaxies. That is, finding a lower luminosity galaxy
near an \ion{O}{6} absorber is less probable than if galaxies were
uncorrelated with the \ion{O}{6} absorbers for $D=1-3h^{-1}$Mpc.
This merely implies that the correlation distance, defined to be the
distance within which there is positive enhancement of pairs compared
to random distributions, is smaller than $1h^{-1}$Mpc for these low
luminosity galaxies, consistent with results summarized in Table 2
(see below). At higher luminosities, the curves are above the random
galaxies. This implies that galaxies with $\gtrsim0.3
L_\mathrm{z,*}$\ are correlated with \ion{O}{6} WHIM absorption with
the correlation distance larger than $3h^{-1}$Mpc, again consistent
with results summarized in Table 2. This same
information may be seen, in a different way, in Figure~\ref{fig:2},
as the families of curves for fainter galaxies lie below equivalent
ones for random galaxies, but the converse is true for higher
luminosity galaxies. A perhaps somewhat counter-intuitive result in
Figure 1 is that an absorber is more likely to find a faint galaxy
within the $D=1h^{-1}$Mpc cylinder, if the galaxies were randomly
distributed, and $\sim 30\%$ of absorbers do not find any galaxy to
the faintest limit simulated. This is due to the fact that the
galaxies themselves are more strongly clustered than absorbers among
themselves or between galaxies and absorbers. Therefore, absorbers
will have a lower probability of finding neighboring galaxies than
when the latter are randomly distributed, beyond the
cross-correlation distance.

An interesting feature immediately visible from Figures~\ref{fig:1}
and \ref{fig:2} is that the association between galaxies and
\ion{O}{6} absorbers depends weakly on either the \ion{O}{6} column
density or $\lambda$1032 equivalent width. This suggests that, while
overall galaxies and \ion{O}{6} absorbers are correlated on small
scales, the physical properties of \ion{O}{6} absorbers (e.g.,
strength, kinematics, number of components) themselves do not
display any tight correspondence with nearby galaxies. It seems that
the strengths of \ion{O}{6} absorbers do not provide useful
indicators for the properties of nearby galaxies. Physically, it
suggests that \ion{O}{6} absorbers may arise in the vicinities of
galaxies in a wide variety of ways through complex feedback and
thermodynamic processes. Complex interactions involving
gravity-induced shocks, feedback and photoionization appear to have
erased or smoothed out any potential trend with respect to
\ion{O}{6} column density or equivalent width. This finding is in
accord with observations \citep[e.g.,][]{p06}.

From Figure~\ref{fig:2}, it appears that the distance out to which
\ion{O}{6} absorbers are correlated with galaxies is a function of
both galaxy luminosity and absorber strength (even though the mere
presence of a galaxy is not tightly correlated with absorber
strength as from Figure~\ref{fig:1}). Comparison of the black curves
in Figure~\ref{fig:2}, showing the results of uncorrelated absorber
and galaxy positions, with the equivalent families of the curves for
the simulated galaxies and absorbers shows that there is typically a
distance beyond which it is more probable to find an uncorrelated
galaxy. This distance changes depending on the galaxy luminosity and
the absorption strength. In the extreme case, there is no distance
at which the strongest absorbers are correlated with the least
luminous galaxies. In Table~\ref{tab:distlum}, we list the
correlation distances as a function of galaxy luminosity and
absorption strength. We note two interesting features from the
table: (1) For a given equivalent width limit, the correlation
distance is a monotonic function of the limiting luminosity. (2)
However, for a given limiting luminosity, the correlation distance
is not a monotonic function of equivalent width limit. This may also
point toward the eclectic nature of the absorbers as mentioned
above.

It is interesting, however, to consider the number of galaxies that
may be responsible for producing \ion{O}{6} absorption. While the
intragroup/intracluster medium (ICM) is typically too hot
($T\sim10^6$\,K) for \ion{O}{6} to survive in appreciable
quantities, the interfaces between warm, denser, photoionized clouds
of temperature $T\sim 10^4$\,K and the ICM are potential locations
for \ion{O}{6} production. Examples of such interfaces include the
boundaries between Milky Way high-velocity clouds and the hot
Galactic corona \citep[e.g.,][]{fox05,sem03}. From
Table~\ref{tab:ngal}, we find that 85\%\ of \ion{O}{6}, regardless
of absorption equivalent width, do not lie near $L_\mathrm{z,*}$\
galaxies. Furthermore, the remaining 15\%\ have at most 3 nearby
$L_\mathrm{z,*}$, comparable the Local Group. This is not surprising
given that $L_\mathrm{z,*}$\ are not common. However, 99\%\ of
absorbers do lie near galaxies of lower luminosity, even if the
presence of those galaxies is not correlated over what is expected
from randomly placed galaxies.

Figure~\ref{fig:2} shows that $\le$20\% of \ion{O}{6} absorbers
should find an $L_\mathrm{z,*}$ galaxy within a projected distance
of $5 \mathrm{h}^{-1}$\,Mpc. Of course, one would not likely
associate \ion{O}{6} absorbers with $L_\mathrm{z,*}$ galaxies at
such large projected separations, since, for example, one would have
already found a nearby $\ge 0.03L_\mathrm{z,*}$ galaxy within
1h$^{-1}$\,Mpc, or $\ge 0.1L_\mathrm{z,*}$\ galaxy at closer
distance with comparable probability. In any case, it seems unlikely
that $L_\mathrm{z,*}$ galaxies at such remote distances are
responsible for creating the \ion{O}{6} absorbers. The rapid rise of
probability in Figure~\ref{fig:2} from $D_\mathrm{min}=0$ to
$D_\mathrm{min}\sim 1 \mathrm{h}^{-1}$\,Mpc from galaxies $\ge
0.03-0.1L_\mathrm{z,*}$ may reflect a ubiquitous physical connection
between \ion{O}{6} absorbers and these relatively small galaxies,
perhaps a result of galactic superwinds being able to transport
metals to a distance of $\le 1 \mathrm{h}^{-1}$\,Mpc from these
galaxies. This, however, does not necessarily exclude larger
galaxies from being able to do the same. The slower rise of
probability in Figure~\ref{fig:2} from $D_\mathrm{min}=0$ to
$D_\mathrm{min}\sim 5 \mathrm{h}^{-1}$\,Mpc for galaxies $\ge
0.3-1L_\mathrm{z,*}$ may be a result of the intrinsic correlation of
large galaxies and small galaxies on these scales. These more
detailed issues will be examined subsequently elsewhere. Our results
appear to be in broad agreement with observations
\citep[e.g.,][]{s06,s07,t06}.

We take a more detailed look at the dependencies of probability on
the \ion{O}{6} column density or equivalent width. Closer
examination of the curves in Figure~\ref{fig:1} (lower-right corner
of both panels) and in Figure~\ref{fig:2} (solid curves, lower-left
corner of lower panel) reveals that the probability of finding a
$\sim L_\mathrm{z,*}$ galaxy is higher for \ion{O}{6} absorbers with
a higher column density equivalent width. In particular,
$N$(\ion{O}{6})$\ge 10^{14}$\,cm$^{-2}$\ or $W_\lambda\ge 100$\,
m\AA\ absorbers deviate noticeably from the weaker absorbers. This
is a reversal of the trends from other parts of those figures. For
smaller galaxies, e.g., $\ge 0.3L_\mathrm{z,*}$ galaxies (the set of
dotted curves in the lower panel of Figure~\ref{fig:2}), the trend
is considerably weaker although still visible. For still smaller
galaxies, the trend is reversed, with weaker absorbers having a
higher probability than stronger ones. These results seem to suggest
that these very strong \ion{O}{6} absorbers tend to be produced in
richer, high density environments where the probability of finding
massive galaxies is enhanced (we will examine this physical link
elsewhere). The fact that weaker \ion{O}{6} absorbers have a higher
probability of finding a galaxy than stronger absorbers, for
galaxies less luminous than $\sim 0.2-0.3L_\mathrm{z,*}$\
(Figure~\ref{fig:1}), once again suggests that these relatively
weaker \ion{O}{6} absorbers ($W_\lambda\le 50$\,m\AA) are probably
produced by galaxies of $\sim 0.1L_\mathrm{z,*}$, not by more
luminous galaxies. This is consistent with the trend in
Figure~\ref{fig:2} above.

\section{Conclusions}

We investigate the relationship between galaxies and metal-line
absorption systems in a large-scale cosmological simulation with
galaxy formation included. Our detailed treatment of metal
enrichment and non-equilibrium calculation of oxygen species allow
us, for the first time, to carry out quantitative calculations of
the cross-correlations between galaxies and \ion{O}{6} absorbers. We
examine the the cross-correlations between \ion{O}{6} absorbers and
galaxies as a function of projection distance, with the
line-of-sight velocity separation between an absorber and a galaxy
constrained to within $\pm 1000$km/s. Here are some major findings:
(1) The cross-correlation strength depends only weakly on the
strength of the absorber but strongly on the luminosity of the
galaxy. This result suggests that \ion{O}{6} absorbers are produced
ubiquitously and their physical/thermal properties and history vary
widely, presumably due to the combined effects of gravitational
shocks, feedback, photo-ionization and cooling processes. (2) The
correlation length, however, does depend on both the galaxy
luminosity and on the absorber strength from $\sim 0.5-1h^{-1}$Mpc
for $0.1L_*$ galaxies to $\sim 3-5h^{-1}$Mpc for $L_*$ galaxies.
While the dependence on luminosity is monotonic, the dependence on
limiting equivalent width appears to peak at some
luminosity-dependent value and then falls. (3) Only 15\%\ of
\ion{O}{6} absorbers lie near $\ge L_\mathrm{z,*}$\ galaxies. Thus
the remaining 85\%\ must be produced by gas ejected from fainter
galaxies. The positions of lower-luminosity galaxies is not well
correlated with absorbers (i.e. in comparison with randomly-placed
galaxies). This may point toward pollution of intracluster gas by
many galaxies, rather than a single high-luminosity galaxy. (4) For
$\gtrsim 0.5L_\mathrm{z,*}$\ galaxies, there is a positive
correlation between absorber strength and galaxy luminosity
(stronger absorbers have a slightly higher probability if finding
such a large galaxy at a given projection distance).  The reverse
seems true for less luminous galaxies.  On average, these results
indicate that very strong \ion{O}{6} absorbers tend to be produced
in richer, high density environments where it is more likely to find
massive galaxies.

The spatial resolution of our simulation is $\sim 80h^{-1}$kpc.
While this is adequate for resolving large galaxies, it becomes
marginal for galaxies in halos of total mass less than $\sim 10^{11}
M_\odot$. Therefore, some of the smaller galaxies of luminosities
$0.01-0.03L_\mathrm{z,*}$ may be significantly affected and their
abundances underestimated. In addition, \ion{O}{6} systems that
would have been produced from these under-resolved galaxies may be
absent. As a result, one should treat the cross-correlation strength
between \ion{O}{6} absorbers and the low-luminosity galaxies as a
lower bound. But we hope that the preliminary results presented here
will provide a useful framework for comparison with upcoming imaging
campaigns of galaxies in the field of quasars and spectroscopic
observations with the {\it Cosmic Origins Spectrograph}.

\acknowledgements{We thank Ken Nagamine for providing simulated
galaxy catalogs, and the referee for thoughtful comments. We
gratefully acknowledge financial support by grants AST-0507521 and
NNG05GK10G. This work was partially supported by the National Center
for Supercomputing Applications under MCA04N012.  }

\end{document}